\documentclass[a4, 11pt]{article}

\usepackage{amsmath}
\usepackage{graphicx}
\usepackage{times}
\usepackage{color}
\usepackage{hyperref}
\usepackage{subfig}
\usepackage{lineno}

\newcommand{\sgn}{\operatorname{sgn}}

\title{Rapidly evaluating lockdown strategies using spectral analysis: the cycles behind new daily COVID-19 cases and what happens after lockdown.}

\author{Guy P. Nason\thanks{Department of Mathematics, Imperial College, London, South Kensington Campus, London SW7 2AZ, UK. {\tt g.nason@imperial.ac.uk}
}}

\begin{document}

\maketitle

\begin{abstract}
Spectral analysis characterises oscillatory time series behaviours such as cycles, but  accurate estimation
requires reasonable numbers of observations\cite{priestley:spectral}.
Current COVID-19 time series\cite{ecdpc:download} for many countries are short: pre- and post-lockdown
series are shorter still.
Accurate
estimation of potentially interesting cycles within such series seems beyond reach.
We solve the problem of obtaining accurate  estimates from short time series by using recent Bayesian spectral fusion methods.\cite{nason:should} 
Here we show that transformed new daily COVID-19 cases for many countries generally contain three cycles operating at wavelengths of around 2.7, 4.1 and 6.7 days (weekly). We show that the shorter cycles are suppressed
 after lockdown.
The pre- and post lockdown differences suggest that the weekly effect is at least partly due to non-epidemic factors, whereas the two shorter cycles seem intrinsic to the epidemic.
Unconstrained, new cases grow exponentially, but the internal cyclic structure causes periodic falls in cases. This suggests that lockdown success might only be indicated by four or more daily falls in cases.
Spectral learning for epidemic time series contributes to the understanding of the epidemic process, helping evaluate interventions and assists with forecasting. Spectral fusion is a general technique that is
able to fuse spectra recorded at different sampling rates, 
which can be applied to a wide range of 
time series from many disciplines.
\end{abstract}

During the UK Government COVID-19 briefing  on 6th April 2020, the UK Deputy's Chief Scientific adviser,
Professor Angela McLean, said\cite{mclean:metro}
``We need a good long time series of data on all stages of infection in order to be able to tell what
the impact of measures that came in on March 23 will be''. The measures that
Professor McLean referred to
 were the widespread UK social distancing and lockdown interventions made in the
face of the COVID-19 threat.
At the time of writing, few countries have experienced in excess of 70 days
of COVID-19 cases and most only have around 50 days. Professor McLean is correct in that many scientific inferences
require longer time series than those currently available.
However, we show that there are considerable and useful
similarities in the underlying cyclic (spectral)  behaviours of the numbers of new daily COVID-19 cases for a range of
different countries (see Extended Data figures).
We  use recent Bayesian spectral fusion methods\cite{nason:should} (regspec) to pool spectral information across
countries, which provides significantly more accurate estimates of cyclic behaviour than provided by
a typical spectral analysis
of a single country alone. The Bayesian principles underlying our fusion method handle mean that
uncertainty is treated coherently, producing
 rational uncertainy assessment  for our cycle (spectral) estimates.
 Our methods produce cycle estimates  using the equivalent of
over nine hundred daily observations, compared to the fifty or so that a typical standard spectral analysis might
use. 
Using data\cite{ecdpc:download} from all of the countries we considered, our 
results show that transformed new daily COVID-19 cases  have three underlying cycles:
one operating at a wavelength of 2.7 days, a second
 at 4.1 days and a third at 6.7 days, which we take to be a weekly effect. We conducted separate analyses
 for the UK and groups of countries with similar spectra and note some variation in those cycles.
 
 
 For some purposes it is not reasonable to compare or pool the number of new daily cases from one
 country to another\cite{roosa:short}.
 For example, different countries might use different definitions of the number of daily
cases and they record cases through different national structures and this is even the case for countries with
political, geographical or cultural similarities. However, as long as the method
of recording cases is broadly unchanged over the period in question for a particular country, the spectral properties
across countries are comparable. The transformed cases'
spectrum quantifies the internal oscillatory structure within
the series and is largely unaffected by the overall level of cases,
the different start times of epidemics in different countries (phase) and country-specific internal delays due to
reporting requirements (also phase). In addition, the demonstration of the presence
three consistent cycles across all countries, with some variation, 
 provides supporting evidence for the suitability of the transformed new daily cases as a target of analysis,
 and comparisons between and  across countries,

 Another topic of great current interest is to ascertain whether and how a lockdown will influence the
number of new daily COVID-19 cases. We consider this question for the group consisting of
the UK, Italy, France, Germany, Spain, Switzerland, Belgium and the Netherlands. The number of days
(with cases)
before lockdown  is, on average, 22 for this group of countries, and,
after lockdown, is 26 (except the UK, which started its lockdown later). The averages just quoted include
allowance
for a seven day incubation period. Our analysis compares the spectral properties before and
after lockdown. A spectrum based on about 25 days worth of data would provide a very poor and highly
uncertain estimate. However,  our spectral fusion methods\cite{nason:should} permit 
effective sample sizes for the group of 192 days worth of data prior to the lockdown, and 196 after, resulting
in highly accurate spectral estimates for these periods. We learn that, after lockdown,
the weekly cycle remains strong, but the cycles operating around 2.7 and 4.1 days become suppressed.
This indicates that  the weekly cycle is due, at least in part, to administrative recording effects, which
are  not effected by the lockdown, whereas the
2.7 and 4.1 day cycles might be related to virus dynamics, which is certainly affected by lockdown. 

The discovery of how the high-frequency  cycles are disrupted by full lockdown suggests
that they could be monitored during partial lockdowns. For example, if schools are reopened and
the 2.7 and 4.1 day cycles do not reappear, then this might indicate the effectiveness of that strategy.
Given the similarity of the cycles across countries, this indicates that cases could be monitored and
pooled across regions, over a short number of days to be fused into longer effective samples using
the methods described here.

A more difficult problem is that of forecasting transformed new daily COVID-19 cases. Such information
would be of great interest, e.g., to those planning health provision over a short timescale. Knowledge of the
three cycles is helpful and we have had moderate success in forecasting daily cases.
However, with individual country series,
with smaller number of days, it is unrealistic to expect too much and, in particular, the transformed cycles
experience both a degree of time-modulation and possible frequency changes.
More useful perhaps, are not daily forecasts, but the knowledge that the number
of cases will increase and decrease over a period of three/four days. This means that if one observes
a decrease in the number of daily COVID-19 cases after lockdown, that does not necessarily mean
the peak has been reached, but is simply a manifestation of the 3/4 day cycles. Hence, one
might believe a lockdown strategy has been successful after a sustained decrease of at least four days.

Spectral analysis\cite{priestley:spectral, percival:spectral} of epidemics is not new,
but most  work has been carried out on epidemics observed
over long time periods (seasons
and years) using lengthy time 
series\cite{grenfell:travelling,conlan:seasonality, ferrari:the}. 
Recent work\cite{benvenuto:application} on COVID-19 has applied popular
autoregressive integrated moving average process\cite{chatfield:the, priestley:spectral}models to a single
prevalence time series  with a sample size of $n=22$. However,  conclusions derived from
such analyses on a single series with such small sample sizes\cite{hyndman:forecasting} are  questionable.
For example, an autoregressive process of order one with parameter $0.9$, normally considered to be 
a strong signal, is only distinguishable from white noise\cite{nason:white}
approximately 20\% of the time with sample size of $n=22$;
basic simulation studies show the large number of possible different models that can fit such
short series apparently well. This indicates that it is virtually impossible to tie down the correct model
with such a small sample size.
Phenomenological sub-epidemic models\cite{chowell:a, roosa:short} show much more promise and have been applied
with some success to short-term forecasting of COVID-19 cases in Guangdong and Zhejiang, China. These
improve performance by using bootstrap methods on  short case time series, but are still ultimately
based on a parametric model of single series.
Our work is very different as it provides exceptionally
accurate spectral estimates for a novel live epidemic that is still in its early days on short series, 
but  reliably so by using recent Bayesian spectral fusion techniques.\cite{nason:should}. The nonparametric
nature of our analysis also permits us to split case time series at a boundary (e.g.\ lockdown or other intervention)
and analyse the two halves separately, still with very short series in each. This is perhaps harder to do with 
classical parametric
models and to maintain consistency between the two halves.
On the other hand, our method relies on good quality case series from different regions, which is again
not always the case for all epidemics.

\section*{Transformed series, the UK spectrum and fusing the world and Europe}
\begin{figure}%
\centering%
\resizebox{\textwidth}{!}{\includegraphics{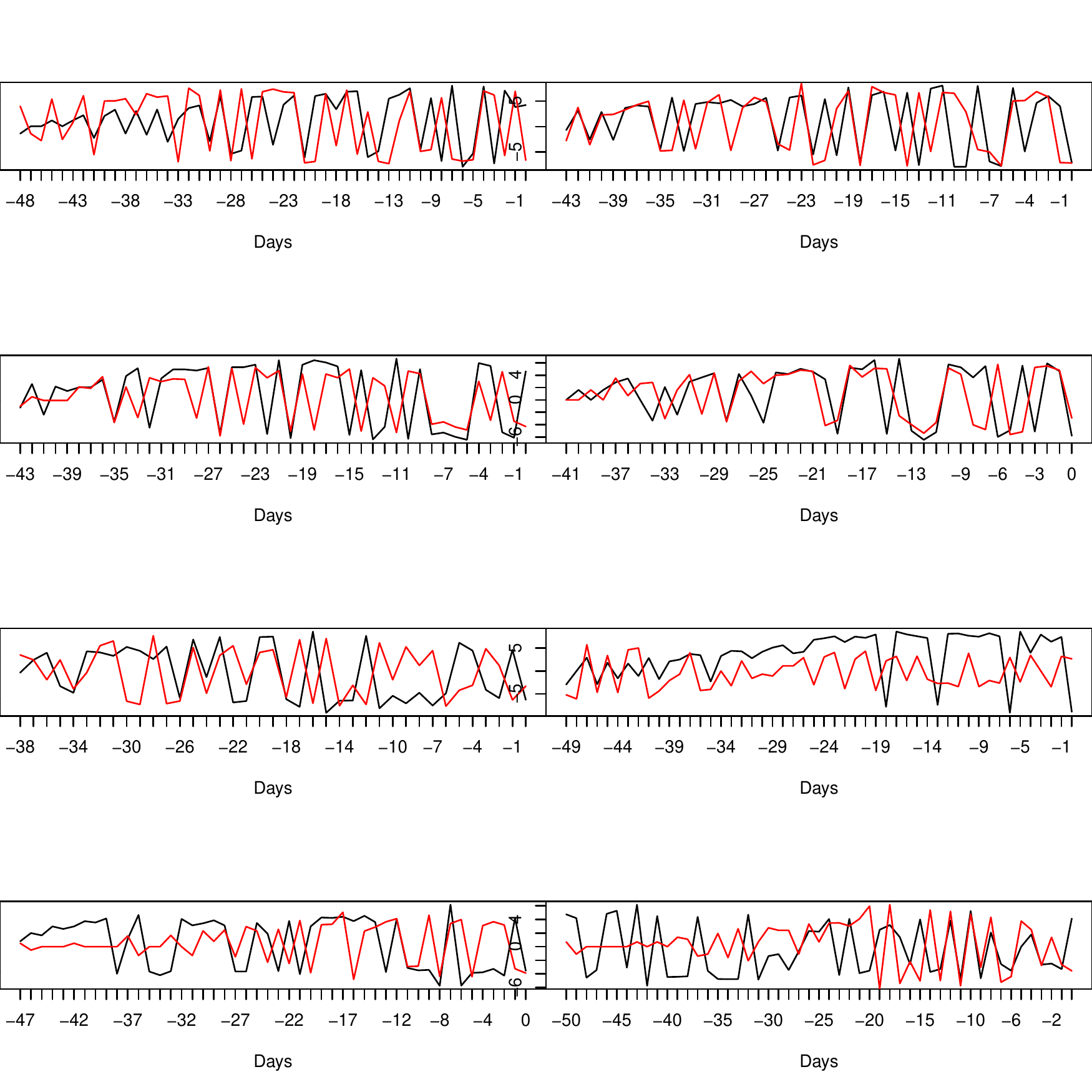}}%
\caption{Number of daily cases on transformed scale for 16 countries. Left-to-right, top-to-bottom:
	UK, IT; FR, DE; ES, CH; BE, NL; AT, NO; US+CN; IR+CA; KR+AU.
	First country of pair in black, second is red.
	\label{fig:transform}%
	}%
\end{figure}

\begin{figure}%
\centering%
\resizebox{0.7\textwidth}{!}{\includegraphics{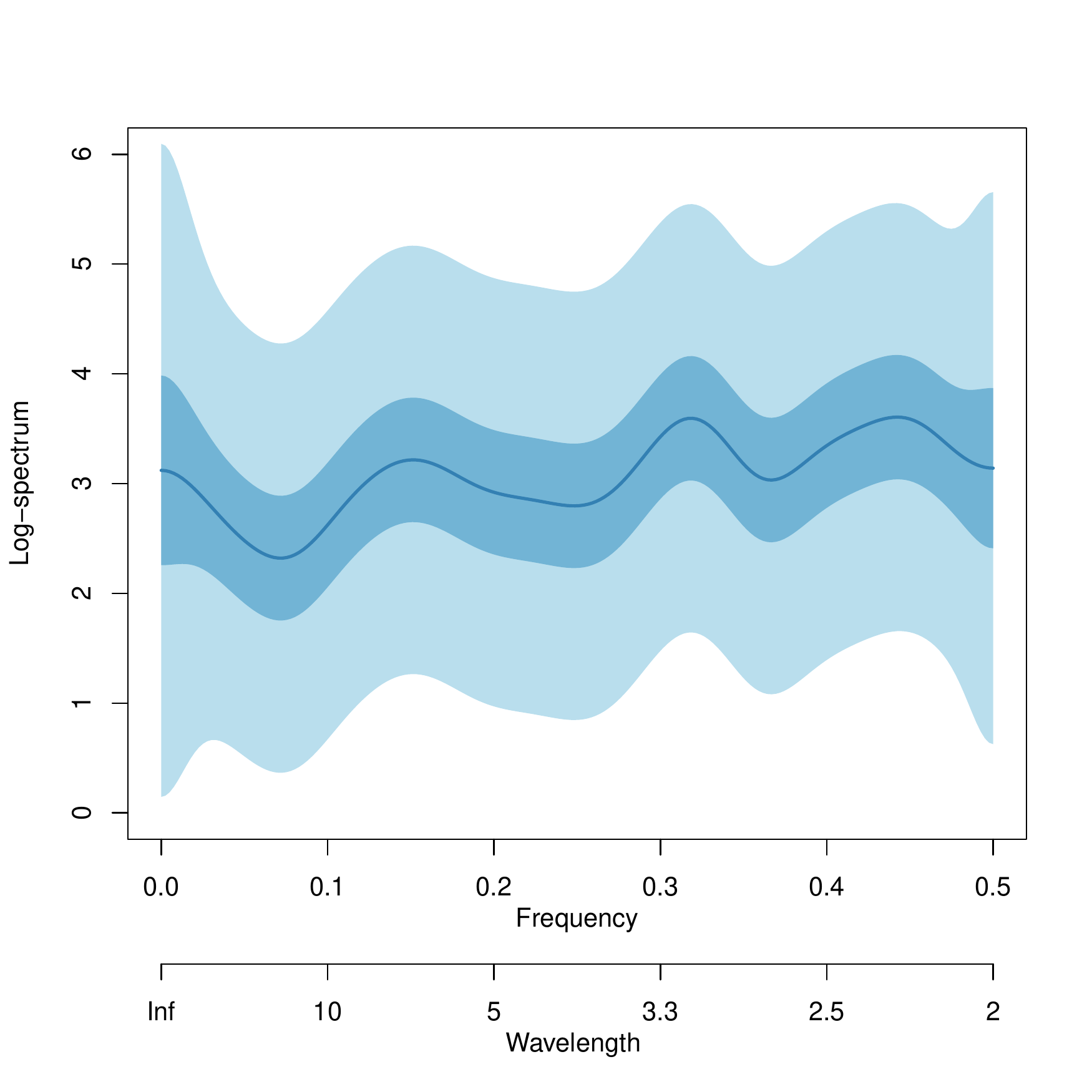}}%
\caption{Bayesian log-spectral estimate of transformed UK new daily COVID-19 cases with
	50\% (dark blue shaded region) and 90\% (light blue shaded region) credible intervals.\label{fig:ukspec}}%
\end{figure}

We transformed the number of new daily COVID-19 cases by applying a signed log transform to
the first differences of the new case time series (see Methods).
The transformed number of new daily cases
for 16 countries are shown in Figure~\ref{fig:transform} each showing
a distorted noisy, but characteristic sinusoidal trace. 
The estimated log-spectrum for the UK transformed new daily cases
is shown in Figure~\ref{fig:ukspec} and for all other countries we analysed in the Extended Data figures.
Spectral estimates are commonly displayed on a logarithmic scale\cite{R}.
Spectral peaks  can be observed
at wavelengths of 6.7, 3.2 and 2.3 days, respectively. Although the peaks are visible, the credible intervals
indicate that there is a fairly large degree of uncertainty,
because this time series contains  52 observations. A frequentist analysis, e.g.\ using
the {\tt spectrum} function in R\cite{R}, produces a similar result, but with  even wider
confidence bands.
Similar spectral analyses for each country indicate three similar spectral
peaks, although not always as well-defined nor in precisely the same location.

\begin{figure}%
 \centering%
\resizebox{0.7\textwidth}{!}{\includegraphics{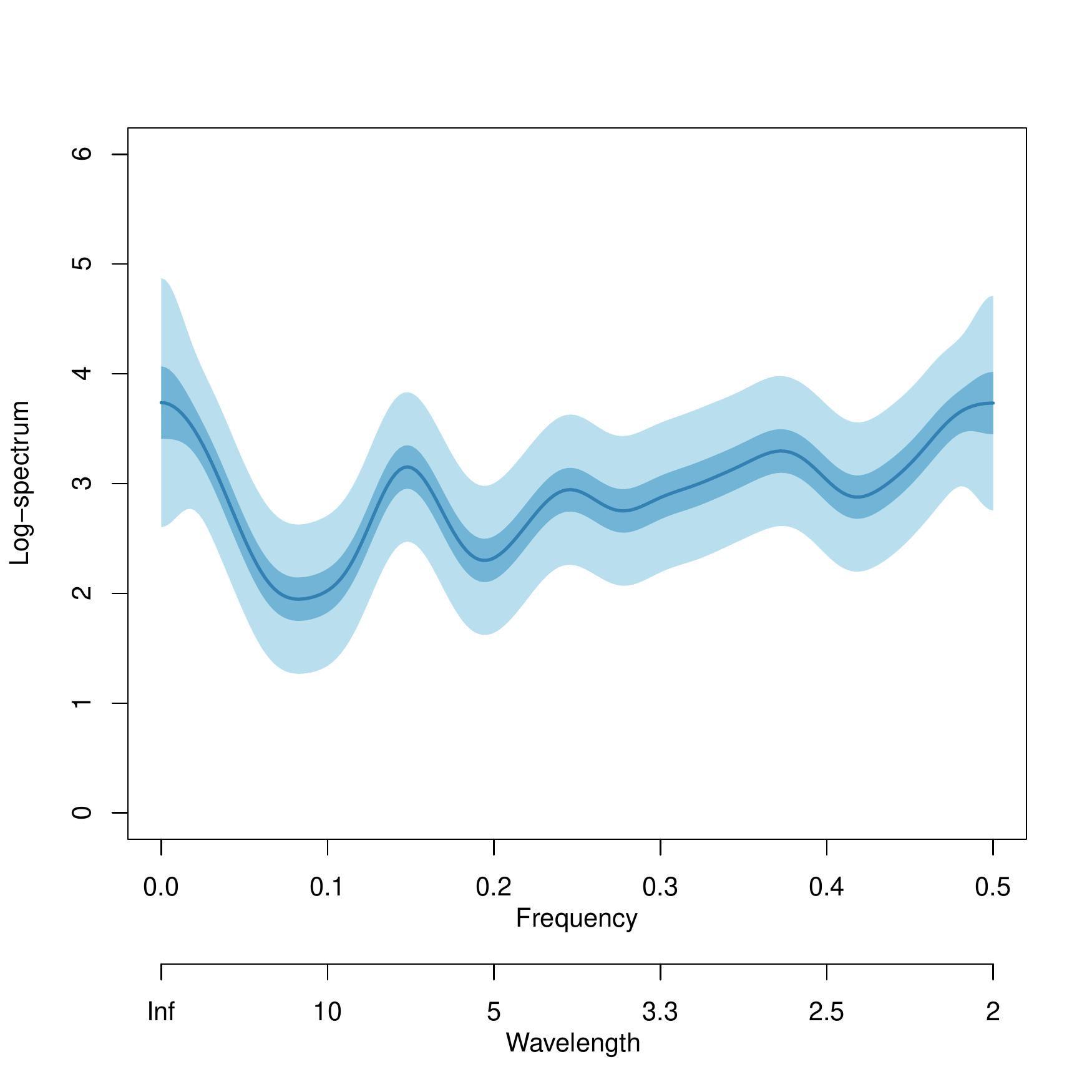}}%
\caption{Bayesian log-spectral estimate of fusion of new daily COVID-19 cases for 18 countries with
	50\% (dark blue) and 90\% (light blue) credible intervals.
	Countries included are the UK, Italy, France, Germany, Spain, Switzerland,
	Belgium, the Netherlands, Austria, Norway, the USA, China, Iran, Canada, South Korea,
	Australia, New Zealand and Sweden.\label{fig:world}}%
\end{figure}

Figure~\ref{fig:world} shows an estimate that is
 the result of coherently fusing  spectra from 18 countries, giving an
an effective sample size of 916 days. Here, the clear spectral peaks have narrow credible intervals, due to the large
effective number of days afforded by using 18 countries together.
The spectral peaks are located
 at wavelengths of 6.7, 4.1 and 2.7 days. The peak around 6.7 days is observed in the spectral plots
 for individual countries and we interpret it to be a weekly effect. Such a weekly 
 effect could be produced by
 reporting artefacts (e.g.\  paperwork being delayed until Monday, or carried out differently at the weekend) 
 or due to the behaviour differences of people at  weekends. All countries analysed have
 a 5+2 working week/weekend pattern, although not necessarily the same days of the week (the actual
 days for a weekend
 are a phase effect, which does not effect the spectrum).

\section*{Clustering spectra and groups of countries with similar  spectra}
\begin{figure}%
\centering%
\resizebox{0.7\textwidth}{!}{\includegraphics{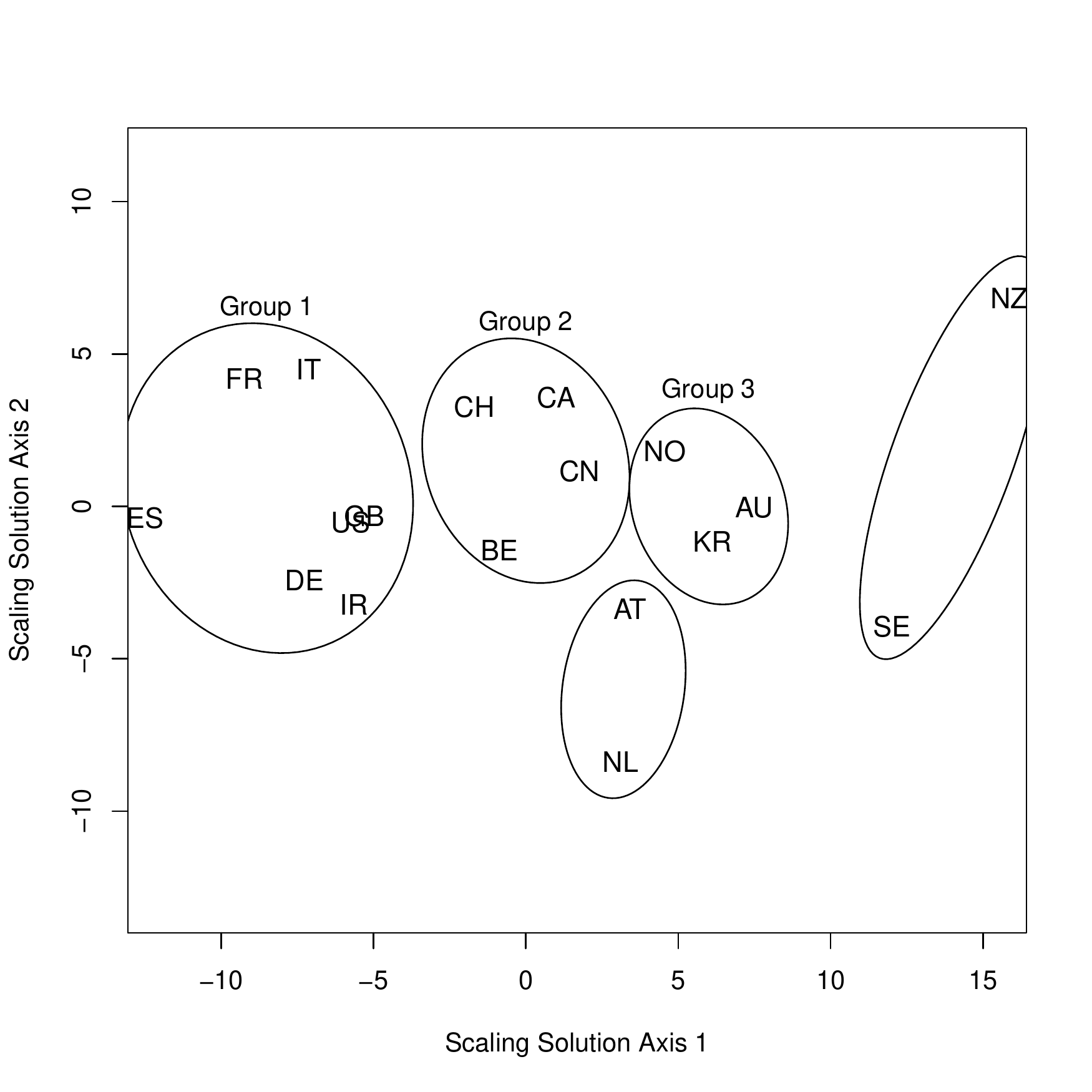}}%
\caption{Multidimensional scaling solution of dissimilarity matrix generated by
	Euclidean distances calculated between two spectral estimates for each pair of
	countries. Countries' position denoted by their two character ISO 3166-1 standard abbreviation.
	The ellipses group together countries that are clustered using the robust
	 cluster stability method\cite{hennig:fpc}.\label{fig:clustscale}}%
\end{figure}

We next clustered our 18 countries
based on their spectrum, by
calculating a dissimilarity between the spectra for each pair
of countries, and then performing both a hierarchical cluster analysis and
multidimensional scaling on the dissimilarity matrix. The scaling solution indicated that
 only two dimensions were required 
to encapsulate 72\% of variation in the data. Figure~\ref{fig:clustscale} shows the resultant two-dimensional
solution.
Attaching a meaning to the scaling axes in Figure~\ref{fig:clustscale} is not easy.
We hypothesise that Axis~1 might indicate how badly a country has been perceived to have been affected by the virus with
Australia, New Zealand and Sweden less so and those on the left of the plot considerably more so. However, Germany
is the obvious anomaly to this interpretation as, currently, it has perhaps been perceived to have handled the crisis well
so far.

\begin{figure}%
 \centering
  \subfloat[\label{fig:g1}]{%
  	\includegraphics[width=0.3\textwidth]{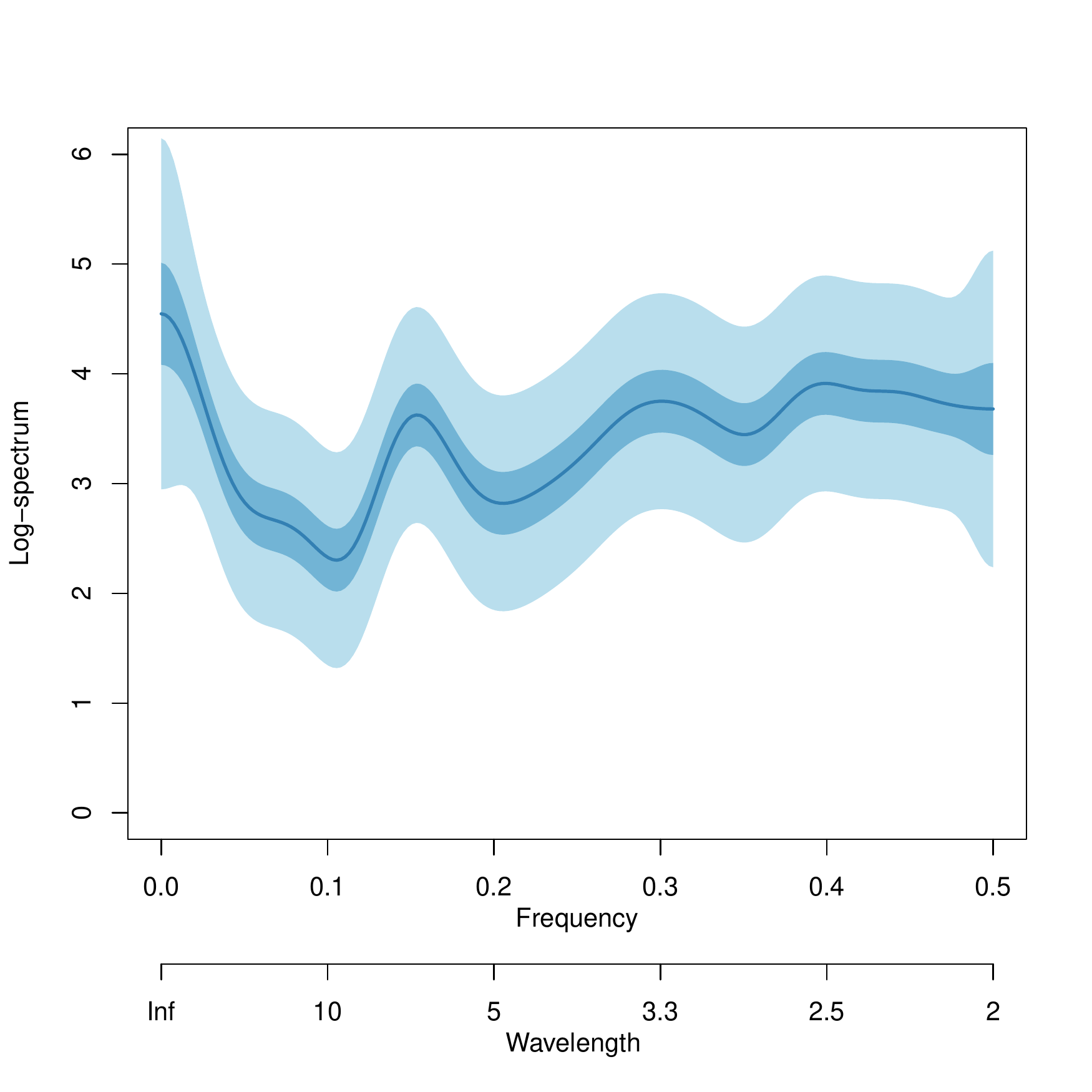}}%
	\hspace{\fill}%
  \subfloat[\label{fig:g2}]{%
  		\includegraphics[width=0.3\textwidth]{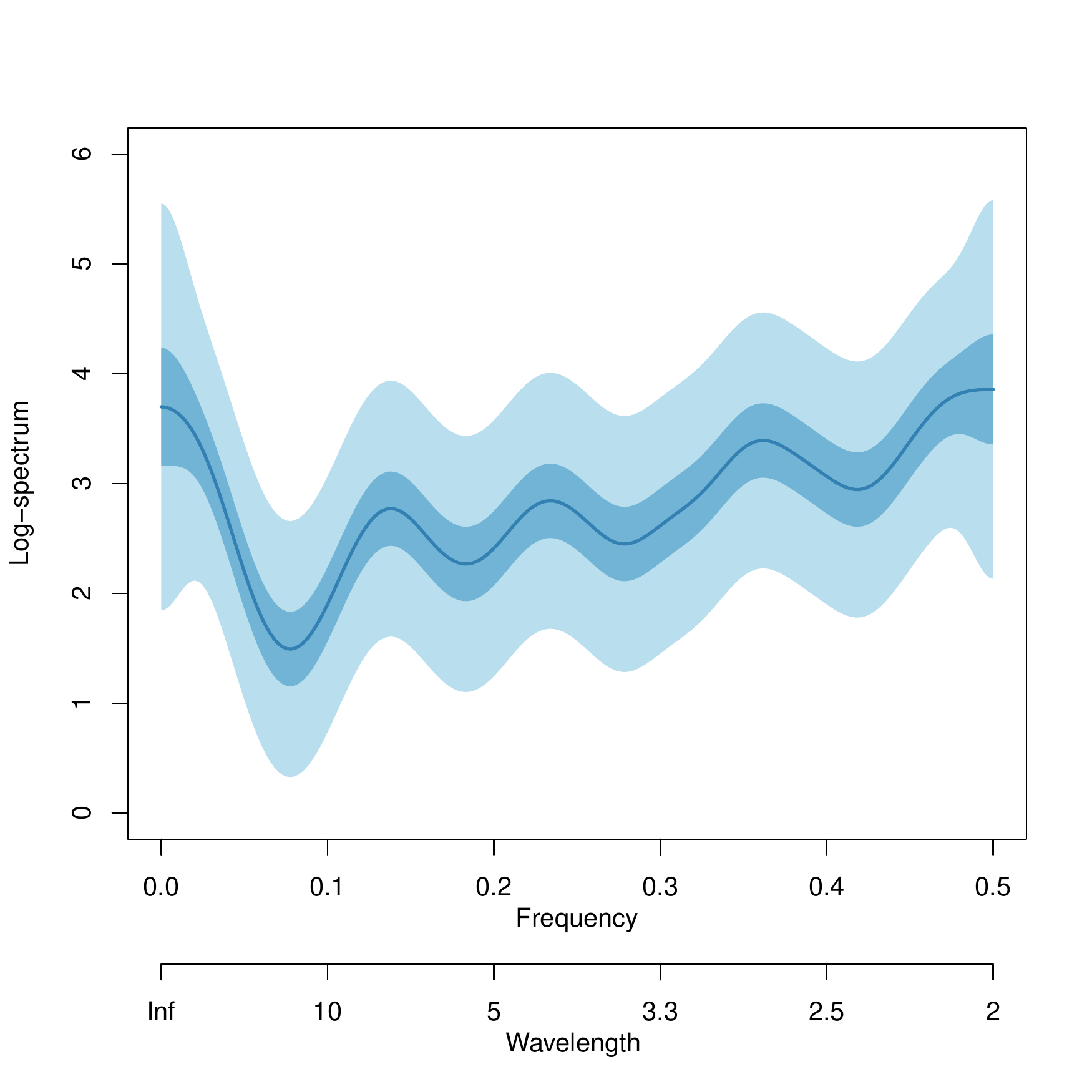}}%
		\hspace{\fill}%
  \subfloat[\label{fig:g2}]{%
  	\includegraphics[width=0.3\textwidth]{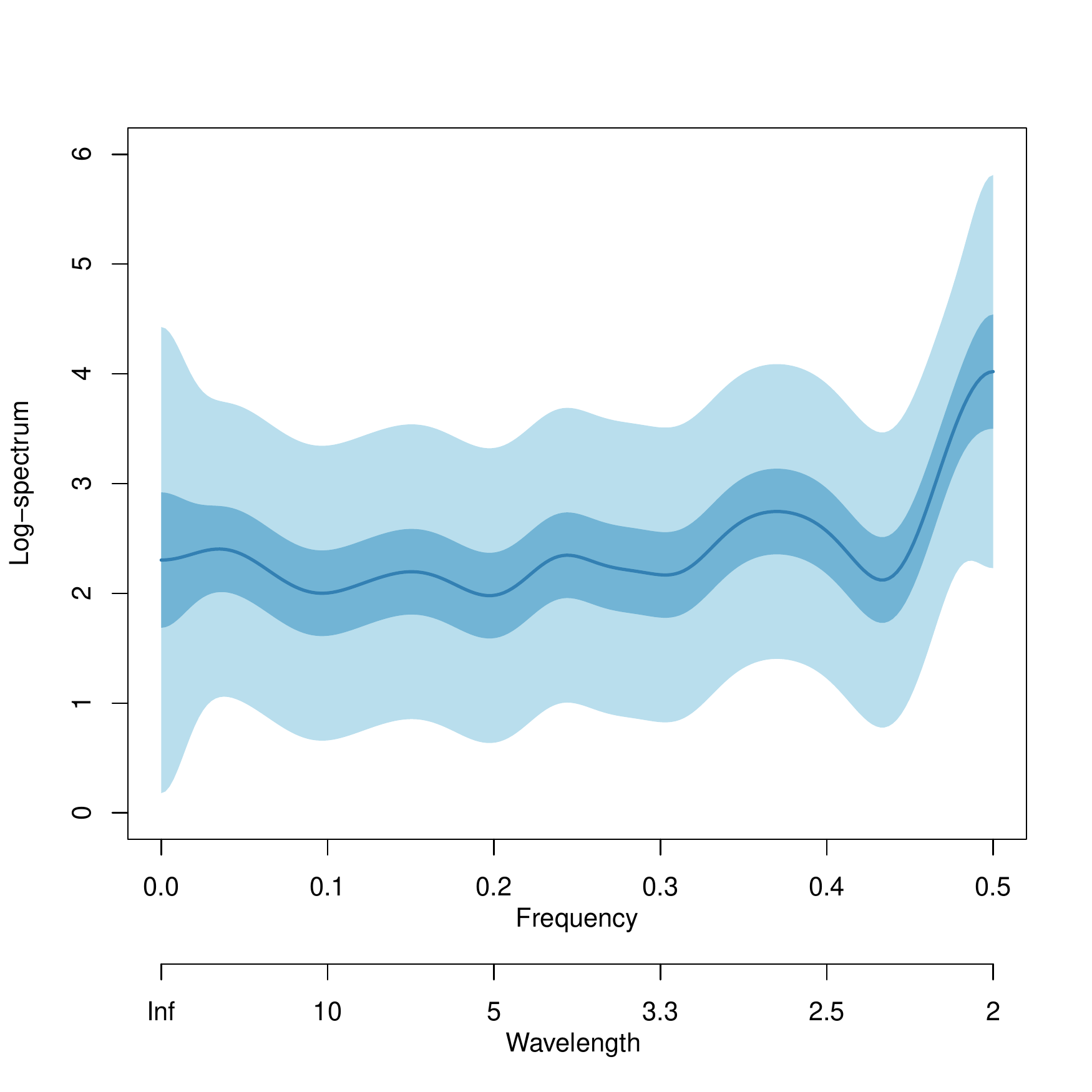}}%
\caption{Bayesian log-spectral estimates and 50\% and 90\% credible intervals for
(a) Group~1 countries: Spain, France, Italy, the US, the UK and Iran.
Effective number of days=357;
(b) Group~2 countries: Switzerland, Canada, Belgium and  China.
Effective number of days=229;
(c) Group~3 countries: Norway, Australia  and South Korea.
Effective number of days=157.%
\label{fig:groups}}%
\end{figure}

Figure~\ref{fig:groups}  show the spectral estimates for the three groups of
countries identified in Figure~\ref{fig:clustscale}, using the clustering techniques mentioned in Methods.
The peak frequencies for each of these groups is listed in Table~\ref{tab:gppeaks}, which shows
differences between them. However, 
each group possesses a possible weekly peak and  higher-frequency peaks labelled
a., of around three to four days, and b., around 2.6 days.
\begin{table}
\centering
\begin{tabular}{c|c|c|c}
Peak & Group 1 & Group 2 & Group 3\\\hline
Weekly & 6.48 & 7.27 & 6.59\\
a. & 3.31 & 4.30 & 4.09\\
b. & 2.52 & 2.77 & 2.70
\end{tabular}
\caption{Spectral peaks for the three country groups in units of days. The peaks in the second
and third rows have been arbitrarily labelled as peak a.\ and b. \label{tab:gppeaks}}
\end{table}

\section*{Spectral changes after lockdown}
Many countries experiencing the COVID-19 pandemic  have instituted a lockdown procedure 
to dramatically reduce virus transmission. At the time of
writing, these countries have observed new daily COVID-19 cases for between 43 and 54 days.
We assume that, on average, it takes about seven days for  the virus to incubate, for
a person to seek attention and then be tested positive for the SARS-CoV-2 coronavirus. For each country,
the number of days prior to and after the lockdown (including incubation time) are
the UK (38, 14), Italy (26,28), France (23, 26), Germany (24, 25), Spain (22, 27),
Switzerland (22, 25), Belgium (18, 25) and the Netherlands (19, 26). 
For some of these countries the lockdown was applied over a period of two of three days and  we took
the median of these as the lockdown start date.

\begin{figure}%
\centering
\resizebox{0.7\textwidth}{!}{\includegraphics{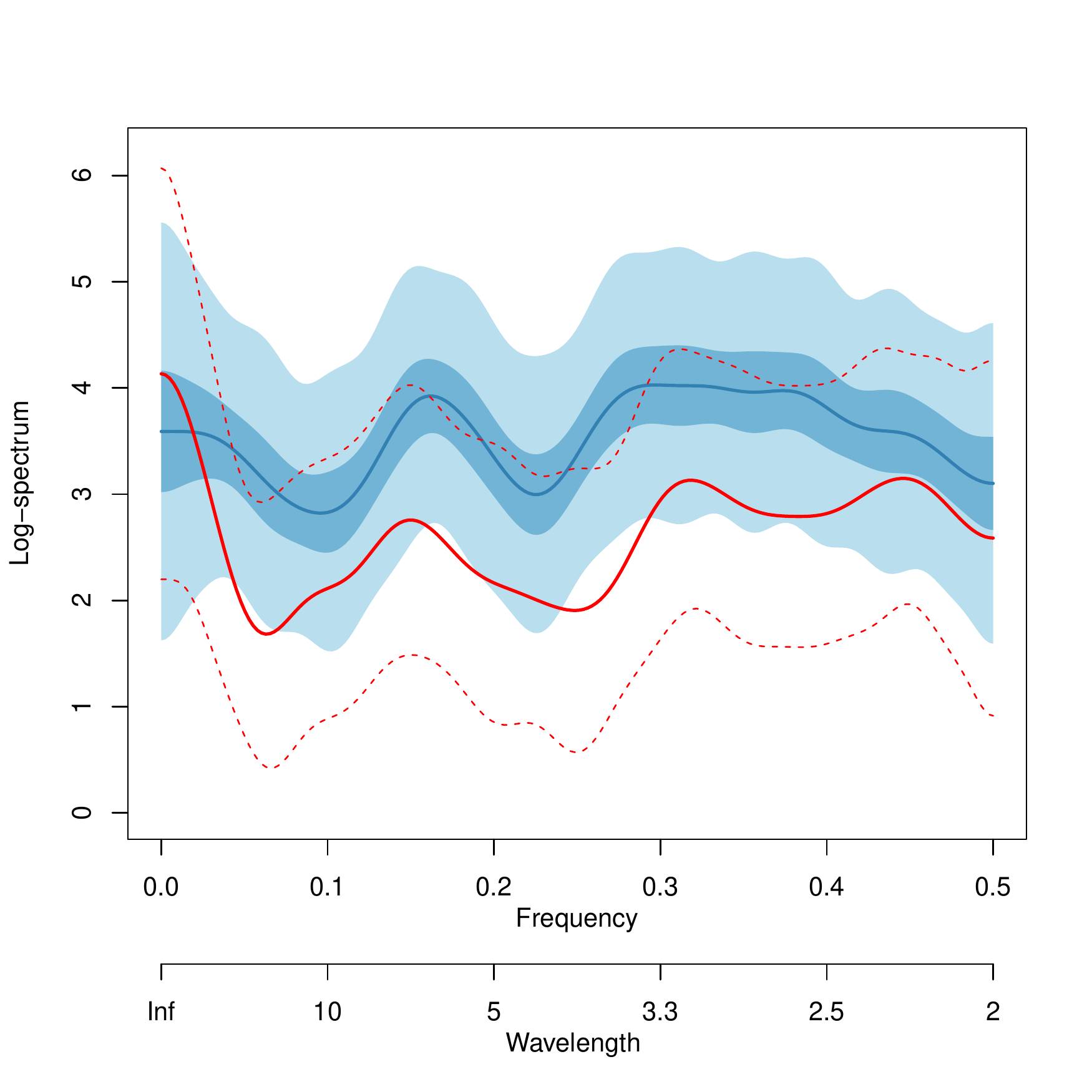}}%
\caption{Pre-lockdown Bayesian log-spectral estimate (red solid line) and 90\% credible interval (red dashed lines).
 Post-lockdown log-spectral estimate (blue solid line) and 50\% and 90\% credible intervals.%
\label{fig:lockdown}}%
\end{figure}

The number of days before and after the lockdown are, in each case, too small to carry out
anything other than the most simplistic time series to maintain statistical reliability.
In particular, a spectral estimate in this situation
would be subject to a high degree of uncertainty.
However, Figure~\ref{fig:lockdown} shows our
coherently fused spectral estimates\cite{nason:should} across these countries before and after the lockdown period, making
use of 192 effective days prior to lockdown and 196 days afterwards.
The weekly peak is clearly
visible in both estimates. The second and third peaks (labelled a.\ and b.\ in Table~\ref{tab:gppeaks}) are
visible pre-lockdown, but have all but disappeared post-lockdown. The spectrum is flat in the location where
peak a.\ was previously, and spectral power declines considerably, relatively,  where peak b.\ was located previously.

This result is particularly interesting as it suggests that peaks a.\ and b.\ have been disrupted by the lockdown.
The weekly effect seems relatively
unchanged by the lockdown, indicating that perhaps it was strongly driven by non-epidemic effects, such as
recording/paperwork or bureaucracy caused by weekends.

The post-lockdown spectrum is higher overall than the pre-lockdown spectrum, this is due to the larger variation
 associated with the larger number of cases identified during the progress of the epidemic. Our 
transformation suppresses this variation, but does not remove it entirely.

\section*{Forecasting daily cases}
We have had varied success in forecasting daily cases using a sum of  two time-modulated cosine waves model,
described in Methods, and more research is required.
We used the Nelder-Mead\cite{nelder:a} optimisation routine built into R\cite{R}, with
starting frequencies of $0.31$ and $0.44$  taken from our UK spectral estimate plots, and
built the model on the transformed cases up to April 11th.
After optimisation, the fitted model resulted in modified frequencies of $\hat{\omega}_1 = 0.34, \hat{\omega}_2 = 0.45$,
close to the starting frequencies
(the other estimated parameters were
$\hat{\alpha}_1 = 1.28, \hat{\alpha}_2 = 0.27, \hat{\phi}_1 = -0.102, \hat{\phi}_2 = -0.074,
\hat{\mu}_1 = 1.3, \hat{\mu_2} = -0.731, \hat{p}_1 = 0.21, \hat{p}_2 = 0.75$).
\begin{figure}%
\centering%
\resizebox{0.7\textwidth}{!}{\includegraphics{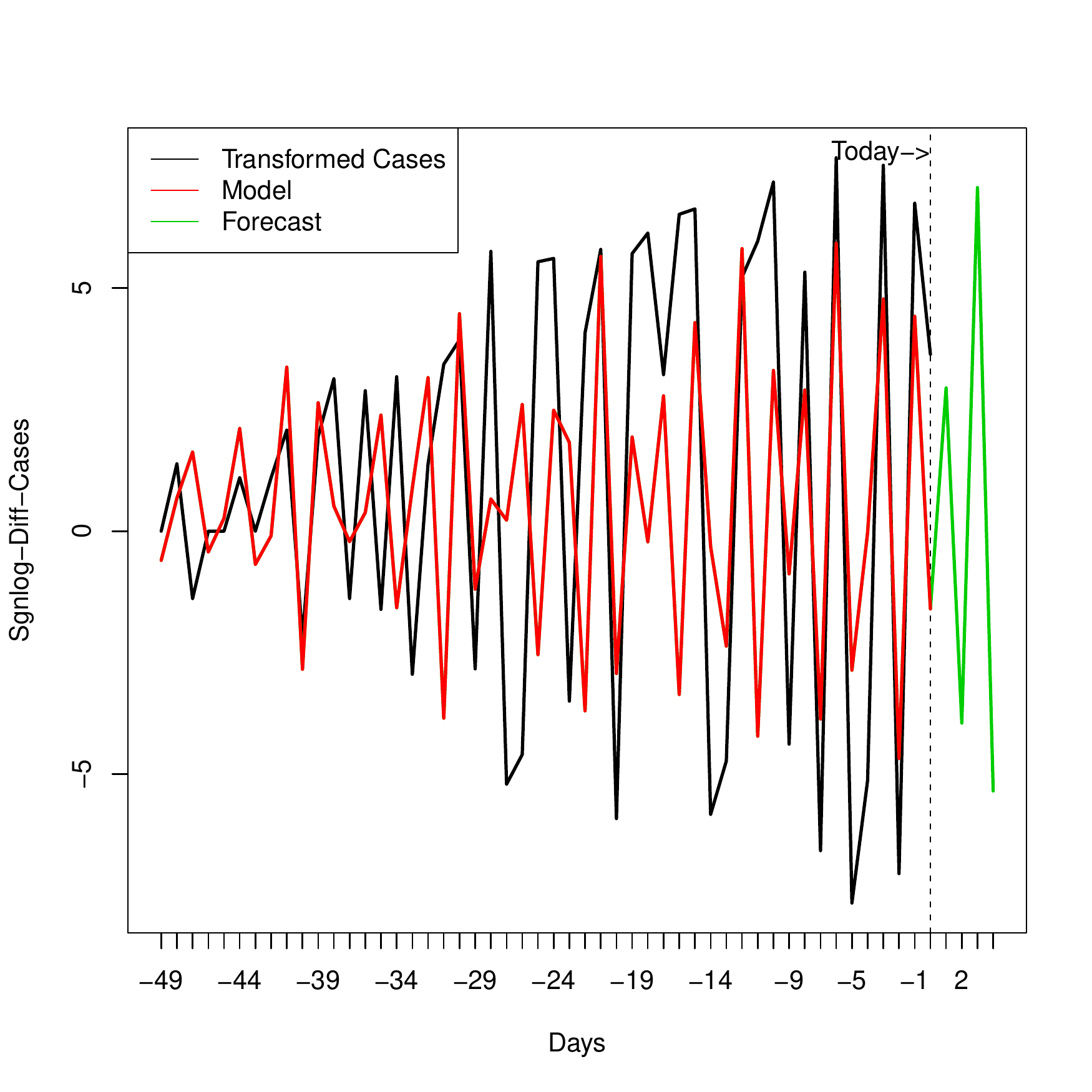}}%
\caption{Number of transformed new UK COVID-19 cases up to and including April 11th (black line),
fitted model (red line) with forecasts (green line). \label{fig:forecast}}%
\end{figure}%
Figure~\ref{fig:forecast} shows transformed new daily UK cases, the model fit and forecasts.
The model fit does not look too bad, many spectral peaks are being identified, but perhaps the amplitudes of
them could be better matched.

The  untransformed
forecasts for April 12th, 13th, 14th and 15th were
5250, 5200, 6373 and 6164, all with approximate 95\% confidence interval of $\pm 150$. The actual number
of cases for April 12th turned out to be 5288. In this case, the forecast was good. However, the two-step ahead
forecast of 5200 was  wrong --- the true value turned out to be 4342 on April 13th.
We also used several stochastic forecasting methods based on autoregressive
integrated moving average modelling and exponential smoothing,
but nothing that we tried was particularly successful. The series is difficult as its amplitude/variance is not constant
and we suspect that frequencies are changing over time (as, e.g., the lockdown plot Figure~\ref{fig:lockdown} indicated).

However, rather than point forecasts, the general sinusoidal nature of the transformed cases suggests a
further, perhaps more reasonable strategy. At this stage, the UK Government and media
are looking expectantly at the daily case numbers to try
and detect a sustained downward trend in cases. Excitement has been generated by a drop in cases two days in a row.
This happened on April 5th with 5903 cases, followed by a drop to 3802 and then 3634 on April 6th and 7th and then,
unfortunately, 
increasing to 5491 on the 8th. However, the general sinusoidal patten, with a wavelengths of about 2.7 and four days shows
that we should only perhaps start believing that a downward turn is a downward trend after a sustained decrease of
four days or more. However,  caution needs to be applied here as  there is no guarantee that the dynamics
will remain unchanged.

\section*{Discussion}

We analysed numbers of deaths using similar methods described here and found similar cycles.
Although we have not carried out a detailed analysis, if the number of deaths
process can be approximated by a linear system\cite{priestley:spectral, chatfield:the} with the numbers of cases as input, then similar cycles are to be
expected.

A time series with a fixed sampling rate and length has a minimum and maximum (Nyquist) frequency that can be
observed.\cite{priestley:spectral, chatfield:the} Although our spectral fusion methods\cite{nason:should} provide
more accurate estimates of the spectrum in the normal range (equivalent to having a larger sample size), they
can not provide information on frequencies outside of the normal range. To estimate lower frequencies,
we would need a genuinely longer series and, for higher frequencies, we would  require cases more
frequently than once a day, which are arguably not really necessary for any practical purpose.

Our analyses  assume approximate stationarity and linearity for the transformed series,
which is unlikely to be exactly true in practice.
For example, in the UK transformed case series in Figure~\ref{fig:forecast}, there are hints of the series
oscillation speeding up over the last ten days. Practically speaking, changes in the testing regime, recording
practices, the lockdowns or
other interventions will change the dynamics of either the pandemic itself or recording of it. Ideally, it would be
of interest to use methods for non-stationary time series\cite{dahlhaus:nonstat,nason:locits}, but the current series available to us are far too short
for such analyses.

\section*{Methods}


All computations were executed in R\cite{R} and packages that are mentioned specifically below.

\subsection*{COVID-19 New Daily Cases Transformation.}

Let $Y_t$, for $t=1, \ldots, n_c$ represent the number of new daily cases for $n_c$ days for country $c$.
The spectral dynamics of the number of daily cases for different countries are all
countries  masked by the well-known and characteristic exponential increases
(and decrease, for those countries that locked
down and have now passed their peak). Hence, we transform our number of daily cases series to
reveal the spectral dynamics. After exploration\cite{chatfield:the} the following transform was used for all series
$L_t = \sgn (D_t)  \log ( | D_t |)$, where the sign function $\sgn(x)$ is $+1$, if $x$ is positive or
$-1$, if $x$ is negative, and $D_{t} = Y_{t } - Y_{t-1}$ for $t=2, \ldots, n_c$. The transform is easily
inverted, which is essential for forecasting the number of daily cases.

\subsection*{Bayesian Spectral Estimation and Fusion: Regspec}
We use the regspec\cite{nason:should,powell:non} Bayesian spectral estimation method with a neutral white noise prior
with prior variance of $1$ and all default arguments, except for  a smoothing parameter of $0.7$, although the results are
not  sensitive to the latter. Regspec straightforwardly enables the production of spectral estimates
using multiple data sets, with each having different lengths and produces coherent credible
intervals to properly ascertain the uncertainty inherent in the estimation process. Regspec can also
 fuse spectra for multiple series recorded at different sampling rates, but we do not need to use
 this aspect of its functionality here as all our case time series are reported daily. However,
 if a country decided to release case numbers on some other sampling plan (e.g.\ every two days, or weekly)
 then Regspec would be able to fuse the spectral estimates as described here. Such a feature
 might be of use when dealing with reporting structures that are not equipped to provide daily reporting of 
 cases or where weekly cases are thought to be more accurate. For example, this might apply
 to regions with fragile health or reporting systems or populations that are spread across widely
 dispersed geographical regions with poor communications.

\subsection*{Clustering of Spectra.} 
Although the number of cases transformed time series show similar spectral behaviour, it is possible 
to observe closer similarities within certain subgroups of countries. We used unsupervised clustering and
scaling techniques\cite{chatfield:introduction,hastie:the} to depict the relationship between different countries
and suggest a clustering for them. First, for each country we produced a spectral estimate using regspec
as mentioned above, and then formed a dissimilarity for each pair of countries by computing the Euclidean
distance between their spectral values
(using the {\tt dist} function in R\cite{R}).
Classical multidimensional scaling was then used to produce an estimated
configuration using the {\tt cmdscale} function in R\cite{R}.
For clustering we use hierarchical cluster analysis on the dissimilarity matrix we computed.
It is well-known that dendrograms are sensitive to the input dissimilarity matrix, so we used the
clusterwise cluster stability assessment by resampling method to produce a stable clustering\cite{hennig:fpc}.

\subsection*{Forecasting}
Given the form of the transformed new daily cases we propose a model, $m_t$, that is the sum of
two time-modulated cosine waves, $m_t = m^{(1)}_t + m^{(2)}_t$, each with formula
\begin{equation}
m^{(i)}_t = \alpha_i \cos \left\{ 2\pi \omega_i (t -\phi_i) \right\} t^p + \mu_i,
\end{equation}
where $i= 1,2$ indexes the two waves and $t=1, \ldots, n_c$.
Initial values for forecast model fitting we used $\alpha_i = 0.8, \phi_i = 0, \mu_i=0.1, p_i = 0.5$, for
$i=1,2$.
For model evaluation we put more weight on getting later observations correct and use 
a residual weight vector $w_t = (t/n)^2$ where $t=1, \ldots, n$ and $n$ is the number of cases.
For short term forecasting, we fit $m(t)$ to the transformed daily cases by weighted least-squares using
standard R\cite{R} optimisation functions and then extrapolate $m(t)$, using recent weighted residuals
to estimate the forecasting error.

\subsection*{Data Availability}
The number of daily COVID-19 cases for countries can be
found at the website of the European Centre for Disease Prevention and Control\cite{ecdpc:download}.



\begin{figure}
\centering
\resizebox{!}{0.75\textheight}{\includegraphics{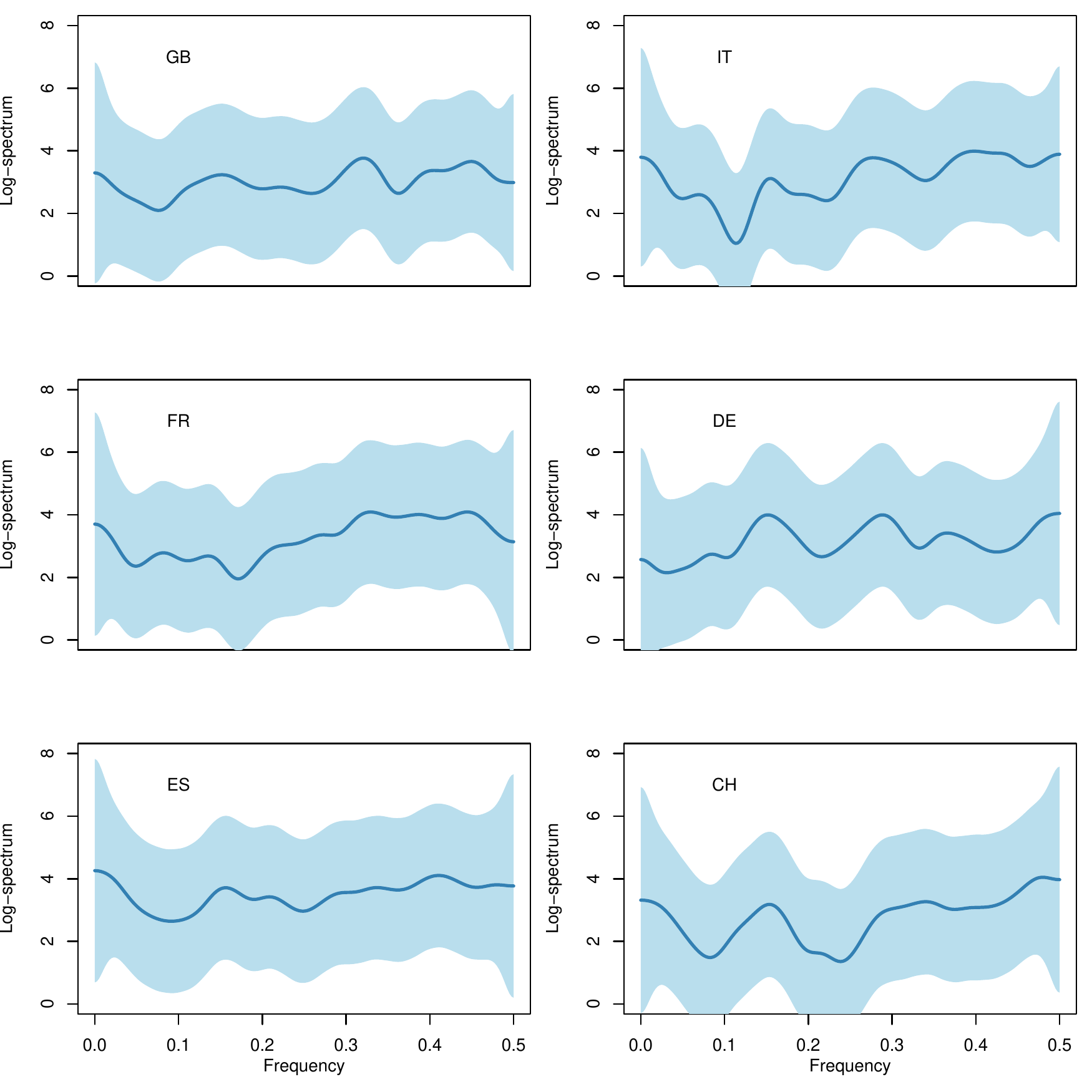}}
\caption{Regspec spectral estimates for numbers of new daily COVID-19 cases. Top-to-bottom, left-to-right:
		United Kingdom on 52 days,  Italy on 54 days, France on 49 days,
		Germany on 49 days, Spain on 49 days,
		Switzerland on 47 days. \label{fig:Ex1}} 
\end{figure}
\begin{figure}
\centering
\resizebox{!}{0.75\textheight}{\includegraphics{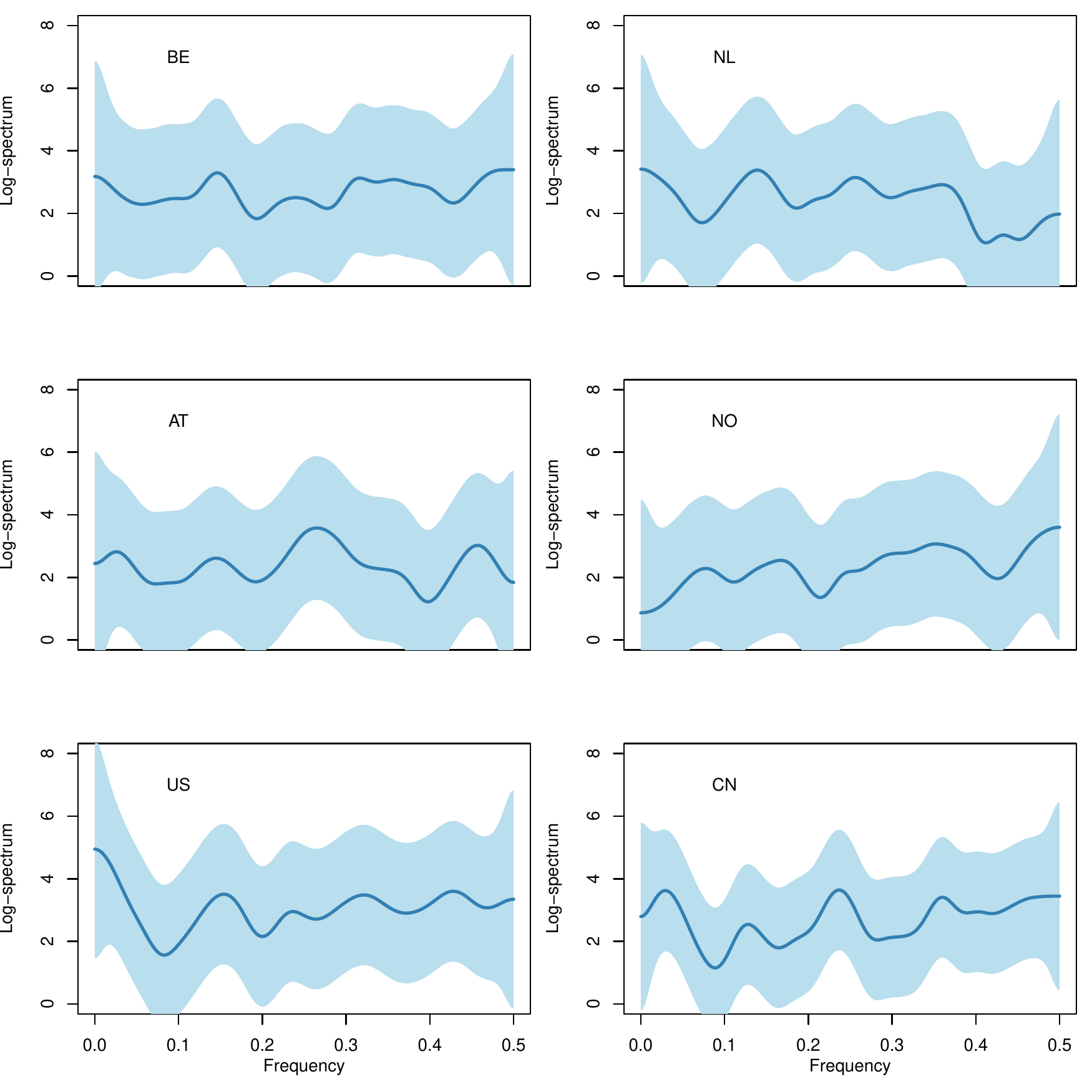}}
\caption{Regspec spectral estimates for numbers of new daily COVID-19 cases. Top-to-bottom, left-to-right:
		Belgium on 43 days,  the Netherlands on 45 days, Austria on 49 days,
		Norway on 47 days, the USA on 53 days,
		China on 87 days. \label{fig:Ex2}} 
\end{figure}
\begin{figure}
\centering
\resizebox{!}{0.75\textheight}{\includegraphics{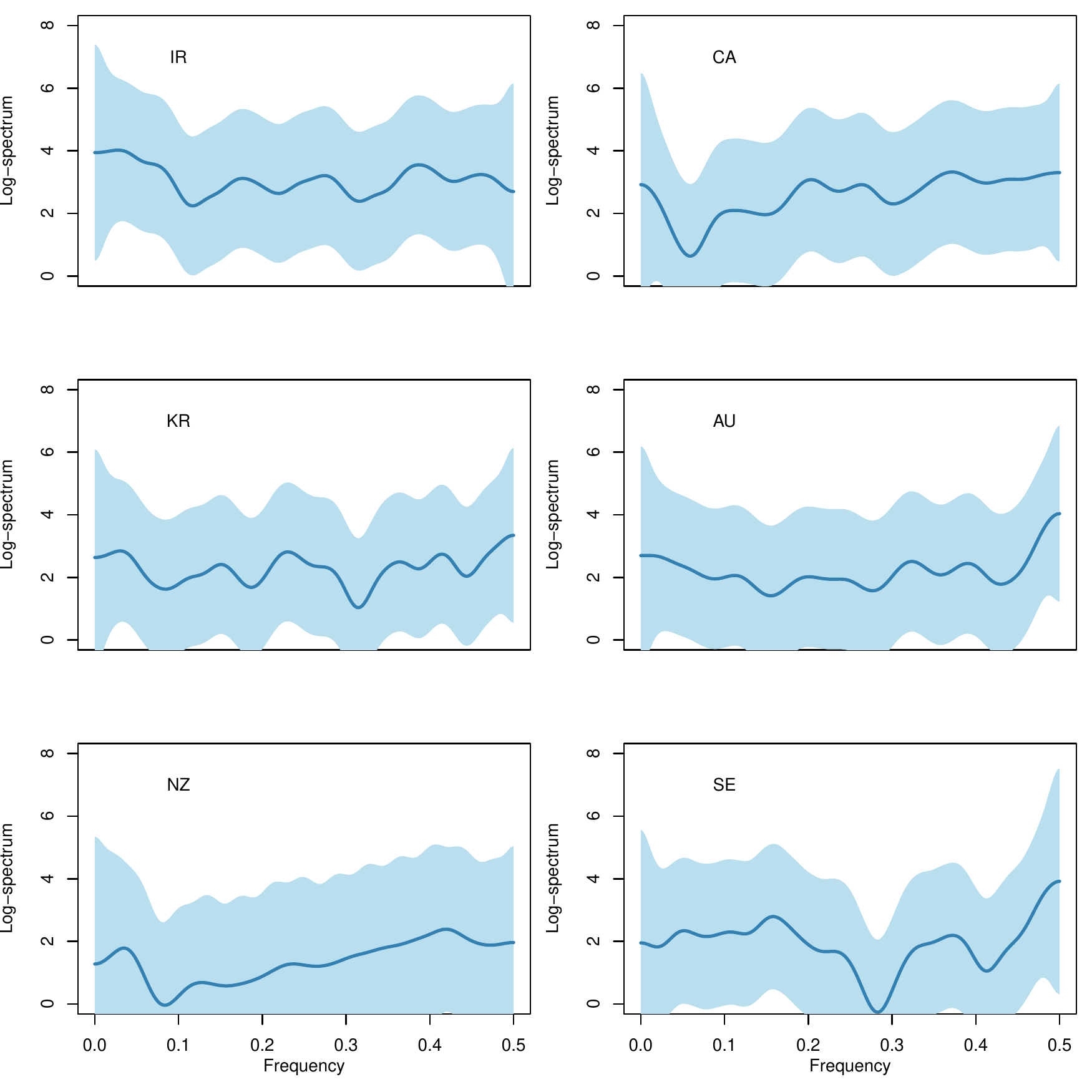}}
\caption{Regspec spectral estimates for numbers of new daily COVID-19 cases. Top-to-bottom, left-to-right:
		Iran on 55 days,  Canada on 49 days, South Korea  on 56 days,
		Australia on 54 days, the New Zealand on 28 days,
		Sweden on 47 days. \label{fig:Ex3}} 
\end{figure}

Extended Data Figures~\ref{fig:Ex1}, \ref{fig:Ex2}, \ref{fig:Ex3} are displayed on the next pages.

\end{document}